\documentclass[conference]{IEEEtran}
\IEEEoverridecommandlockouts
\usepackage{cite}
\usepackage{amsmath,amssymb,amsfonts}
\usepackage{booktabs}
\usepackage{url}
\usepackage{algorithmic}
\usepackage{graphicx}
\usepackage{textcomp}
\usepackage{comment}
\usepackage{multirow}
\usepackage{pdflscape} 

\usepackage{xcolor}
\def\BibTeX{{\rm B\kern-.05em{\sc i\kern-.025em b}\kern-.08em
		T\kern-.1667em\lower.7ex\hbox{E}\kern-.125emX}}

\usepackage{fancyhdr}
\fancypagestyle{FirstPage}

\title{Exploring and optimizing the Vivado Flow on large FPGA designs: A Practical Analysis of Strategies and Their Impact on Timing}

\author{
	\IEEEauthorblockN{KACHKACHI Nor Eddine\IEEEauthorrefmark{1}}
	
	\IEEEauthorrefmark{1}Université Clermont-Auvergne, CNRS, LPCA, 63170 Aubière, \\

}

\begin{document}
	\maketitle
	
	\thispagestyle{FirstPage} 

	\begin{abstract}

		Achieving timing closure in FPGA designs is crucial, especially for high-performance applications. The Vivado Design Suite offers several strategies for optimization, but finding the best configuration for a given project remains challenging. This paper explores the various Vivado’s timing directives and strategies to improve performance across timing, area, and runtime. Through two test-case designs, an NQR spectrometer SoC-FPGA and a digital readout ASIC prototype for high energy physics particle detector's, different strategies and configurations are compared to identify the most efficient flow.
		
	\end{abstract}
	
	
	\section {Introduction}
	\noindent

	Large System-on-Chip FPGA (SoC-FPGA) designs are crucial for high-end applications, but achieving timing closure, managing area, and controlling power consumption is challenging. These designs often consist of programmable logic and hard macro Application-Specific Integrated Circuit (ASIC) sections, leading to complex timing paths. The programmable logic (PL) part may include large and timing-critical modules. Finding the best implementation flow customization with industry-standard tools like Vivado is crucial. The goal of the paper is to explore and compare various Vivado strategies, directives, and tool configurations for synthesis and implementation and show their impact on timing, power and area.

	\section{Background and related Work }
	
	\subsection{Vivado Overview}
	AMD Vivado is a powerful FPGA implementation tool based on the UltraFast Design Methodology (UFDM). It supports simulation, modeling, digital implementation, and embedded software development. Vivado aims to optimize FPGA design performance, area, power, and runtime to shorten time to market. It provides rich strategies, directives, and configurations.
	
	\subsection{Existing Research/Work}
	
	Previous work has addressed Vivado design flow, including general methodologies, comparisons of synthesis and implementation strategies, and timing closure techniques \cite{Monmasson,Lopez}. Resources like UG949, UG904, UG1292 from AMD provide approaches to tackle design challenges \cite{Daughtry}, \cite{AMD_docs}. Open-source Vivado flow scripts are also available\cite{Redpitaya, P.Deminwebsite}.
	
	However, a practical study showing a complete evaluation of all the different Vivado strategies and operating modes and their impact on timing closure on real design test-cases is missing. 
	The objective of this paper is to check first if results obtained
	in the previous study \cite{Lopez} are also valid for the current designs, since in \cite{Lopez} only a small test-case was used.
	Moreover, this paper aims to provide a precise guide to maximize the benefit of available Vivado options.

	\subsection{Challenges in FPGA Timing Closure}
	
\begin{figure}
		\centering
		\includegraphics[width=1\linewidth]{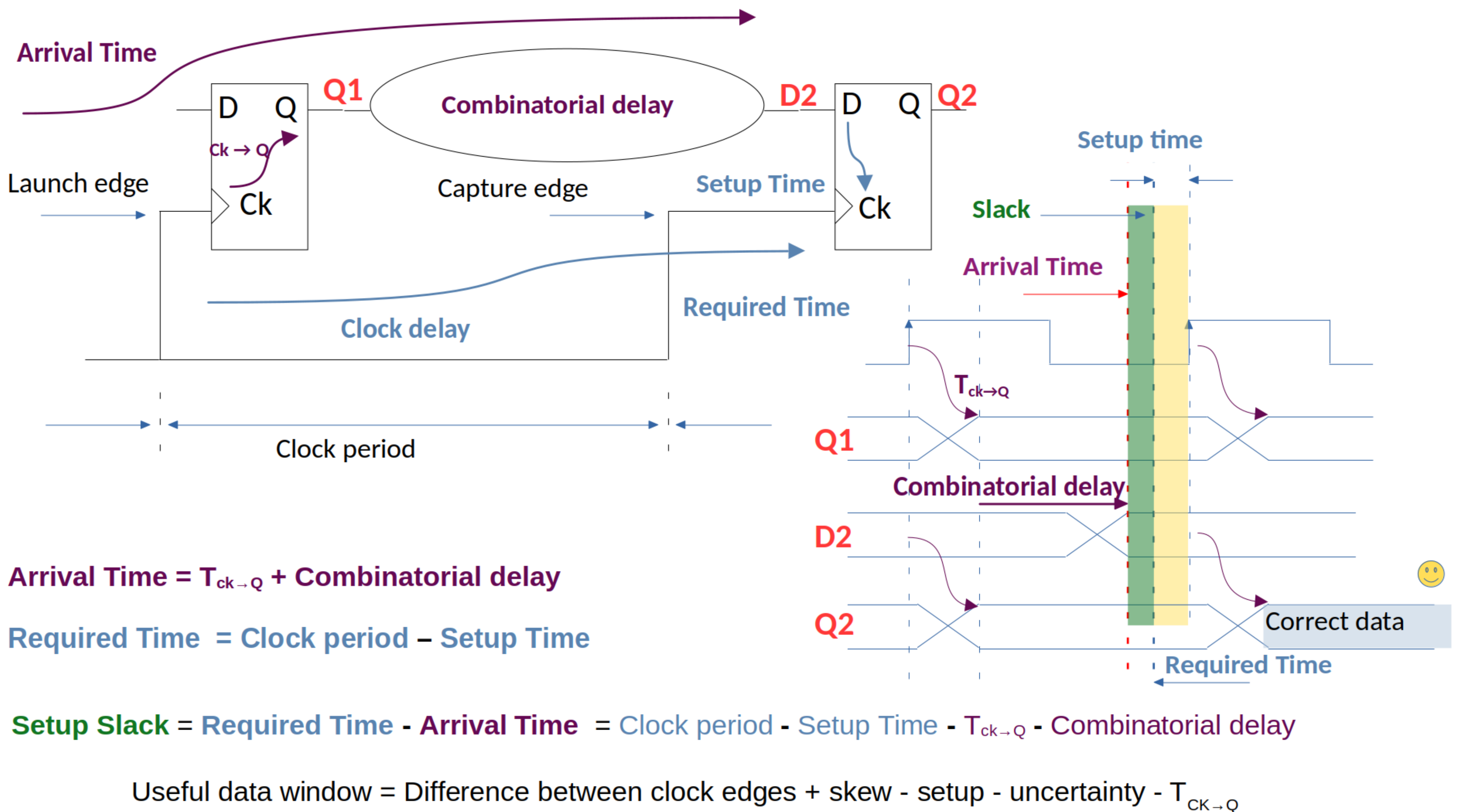}
		\caption{Setup slack check}
		\label{fig:Setup_slack}
	\end{figure}

	Timing convergence is critical in digital design, affecting both ASIC and FPGA performance. Accurate timing analysis is required during the implementation phase. Key timing concepts include setup slack, source, destination, and clock skew. Optimizing the data path is essential for improved performance. Clock skew, clock uncertainty, and Clock Pessimism Removal (CPR) must be considered (figure \ref{fig:Setup_slack}). The UltraFast Design Methodology (UFDM) helps achieve timing closure more quickly through progressive analysis. In this study, the setup Worst Negative Slack (WNS) will be the principal metric to assess timing closure.

	\section{Methodology}

	\subsection{Vivado Design Flow}
	
	The Vivado UFDM Methodology emphasizes iterative analysis using reports to identify and resolve design issues. It involves setting realistic clock frequency constraints and analyzing timing paths. It features baselining which consists in starting with simple timing constraints for internal paths and clock-domain crossings, and add I/O constraints and timing exceptions in later stages. Timing violations should be fixed before moving onto the next step.

	When performing design implementation with Vivado, key concepts are handled.
	Vivado supports project and non project modes. 
	In project mode, Vivado automatically manages the flow, organizes and saves the structure on disk.
	In non-project mode the design is created in volatile memory. The user has full control of the design flow with native Vivado commands.
	
	Strategies are configurations of Vivado that are used in project mode in order to generate reporting, synthesis, and implementation in a predefined way. The users has to possibility to adjust these strategies in order to customize the flow. 
	
	Directives adjust the behavior of the following implementation commands opt\_design place\_design phys\_opt\_design route\_design. 
	
	Design closure means meeting timing and power requirements in conformity with the design rules. It consists in timing closure, power closure and DRC closure.

	The Vivado  flow is based  on several stages: 
	\begin{itemize}
		
		\item{Design entry and checking:} consists in feeding Vivado with the design source and constraints files and then checking for consistency.
		
		\item{Synthesis:} performs sources translation to a netlist, with a crucial impact on timing closure. Synthesis should be iterative and incremental, starting with simple constraints essentially internal clocks. Scientific literature addressed in depth this key FPGA design step \cite{amaru_logic_2017,liu_logic_nodate,mishchenko_scalable_nodate,cortadella_rtl_2015}.
		
		\item{opt\_design:} performs netlist optimization before placement.
		
		\item{Pre-place power\_opt\_design:} an optional phase that performs the optimization of power.
		\item{place\_design:} Consists in a timing and congestion  aware placing the blocks and logic cells. 
		
		\item{Post-place power\_opt\_design:}
		Post placement power optimization phase.
		\item{Post-place phys\_opt\_design:} Post place timing-driven physical optimization.
		
		\item{Back-annotation of phys\_opt\_design:}
		consists in applying post-place optimizations just before placement for subsequent runs.
		
		\item{route\_design:} performs a timing driven routing.
		
		\item{Post-route phys\_opt\_design: } post route timing-driven physical optimization. 
		\item{write\_bitstream:} Generation of the FPGA configuration binary.
		
		\item{Reporting and design analysis:}
		after each step, reporting is performed to check design status before proceeding.
		
	\end{itemize}
	
	\subsection{ Application test-cases and adopted approach}
	
	The designs that constitutes the test-cases gave important challenges.
	A progressive and iterative approach was adopted. An exploration process was followed to compare different synthesis and implementation strategies. TCL scripts were developed and used to launch the different Vivado runs.
	
	\subsubsection {Application test-cases} 
	The current test case consists in the digital readout of an new generation of
	ASIC used for the HEP (High Energy Physics) \cite{eddine_architecture_2026}. It is a matrix of 1024 pixels and an output periphery using a scalable digital architecture that achieves high hit rate with minimal data loss. The digital readout is composed of 256 clusters of 4 pixels each, organized into 16 columns of 16 clusters. It contains several modules, each module organized in control unit (FSM) and an executive unit (data path). The major modules of the ASIC are, a set of 256 digital cluster and a central digital periphery. Each digital cluster manages four pixels. A cluster is composed of 4 pixel modules and a shared arbiter sub-block. Each pixel contains an independent FSM and the shared cluster contains an FSM that manages pixels arbitration and data sending to the digital periphery. The digital periphery contains three major sub-blocks that are, a multiplexer hub sub-block, an address decoder and read-matrix handler. This ASIC was prototyped on a xc7k325tffg900-2 AMD FPGA located on the xc7k325tffg900-2 target using Vivado 2025.2.
	After RTL optimization, some area issues were resolved and the design is
	synthesized and fits well on the FPGA with 18\% LUT and 6\% registers usage. The results obtained so far consist in an implemented digital architecture of the ASIC digital readout on the FPGA with a frequency more than 148 MHz for a target ASIC frequency of 40 MHz. In order to check the optimal implementation strategies for achieving timing closure, the design was over constrained with a frequency of 160\,MHz.
	
	A previous case study was performed in 2024. The application consisted in a large SoC-FPGA design for an NQR spectrometer, based on a AMD Zynq 7010 on the STEMLab14-125 board. A detailed description of this SoC-FPGA is available in the paper \cite{kachkachi_design_2024}.
	The design included an acquisition module, a pulse programmer module, and an excitation module. It uses multiple FIFOs, DMAs, and other IP blocks. 
	
	This design was critical in terms of logic resources which caused severe congestion since the occupancy rate of the Zynq 7010 was almost 100\%. 
	Timing was critical on some internal paths that didn't meet timing for the system clock ($160~MHz$). Due to the real time nature of the application the system clock frequency could not be relaxed in order to prevent data loss in the FIFOs.

	\subsubsection{Adoption of the UFDM as approach}
	
	The UFDM was used to identify timing bottlenecks, analyze and reduce timing violations, resolve CDC issues, reduce compilation times, manage timing exceptions, and use physical synthesis and Retro-fitting.
	For both test-cases, some design updates were performed to improve timing like adding pipelining at the RTL level, recoding adders and multipliers to infer DSP. 
	
	Constraints were defined with timing exceptions on CDC interfaces of the FIFOs since data integrity on these paths is guaranteed by construction. 
	Physical optimization was used after placement and after routing. It helped achieve better timing results.
	
	An incremental flow was used in order to apply post-synthesis and post-route physical optimizations, before synthesis and before placement respectively, in order to improve runtime and implementation results. 
	
	Retro-fitting flow was also used in order to enhance placement and overall QoR. 
	It consists in applying post-placement physical optimizations to the pre-placement netlist.

	Using the Vivado tool, several combinations of synthesis and implementation pre-configured strategies were tested and analyzed.
	Comparison of the results were realized in order to identify the best settings.

	\section{Results}
	\label{Res}

	\subsubsection{Test Setup}
	
	Vivado 2025.2 was used in project mode. 260 combinations between synthesis and implementation runs were launched. For each parent synthesis run, synth\_j, 26 implementation runs denoted as impl\_jxy where tested, covering all the possible pre-built implementation strategies. The index j is the synthesis run number, xy is the implementation run number. The j index varies from 0 to 9 covering the 10 pre-built synthesis strategies that are: Default, Flow\_AreaOptimized\_high, Flow\_PerfOptimized\_high,
	Flow\_AreaOptimized\_medium, Flow\_AreaMultThresholdDSP, Flow\_RuntimeOptimized, Flow\_AlternateRoutability and Flow\_PerfThresholdCarry. These 260 runs include eight runs (table \ref{tab:Run_configurations} ) of the best combinations identified in \cite{Lopez}.

	\begin{table}[ht]
		\renewcommand{\arraystretch}{1.3}
		\caption{\cite{Lopez} Run configurations} 
		\label{tab:Run_configurations}
		\centering
		{
			
			\begin{tabular}{|l|l|l|}	
				\hline	
				Run&Synthesis strategy &Implementation strategy\\
				\hline	
				
				impl\_81&Flow\_AlternateRoutability &Performance\_ExtraTimingOpt \\
				\hline	
				impl\_82&Flow\_AlternateRoutability &Flow\_RunPhysOpt\\
				\hline	
				impl\_83&Flow\_AlternateRoutability &Flow\_RunPostRoutePhysOpt\\
				\hline	
				impl\_84&Flow\_AlternateRoutability &Performance\_NetDelay\_low\\
				\hline	
				impl\_91&Flow\_PerfThresholdCarry &Performance\_ExtraTimingOpt\\
				\hline
				impl\_92&Flow\_PerfThresholdCarry &Flow\_RunPhysOpt \\
				\hline	
				impl\_93&Flow\_PerfThresholdCarry &Flow\_RunPostRoutePhysOpt\\
				\hline	
				impl\_94&Flow\_PerfThresholdCarry &Performance\_NetDelay\_low\\
				\hline	
		\end{tabular}}
	\end{table}

	\begin{table*}[!ht]
		\centering
		\caption{Best Implementation Run Strategy for Each Synthesis Strategy}
		\label{tab:Best_Implementation_Runs}
		\begin{tabular}{|l|l|l|l|l|l|l|l|}
			\hline
			Parent Synth & Implementation Run & Synthesis Strategy & Implementation Strategy & WNS & LUT & FF & ELAPSED \\ \hline
			
			synth\_0 & ~ & Vivado Synthesis Defaults & ~ & ~ & 32517 & 23867 & 00:07:46 \\ \hline
			~ & impl\_006 & ~ & Performance\_Retiming & -1,12 & 31663 & 23913 & 00:10:46 \\ \hline
			
			synth\_1 & ~ & Flow\_AreaOptimized\_high & ~ & ~ & 27642 & 23117 & 00:25:49 \\ \hline
			~ & impl\_101 & ~ & Performance\_ExplorePostRoutePhysOpt & -3,25 & 28267 & 23158 & 00:21:04 \\ \hline
			
			synth\_2 & ~ & Flow\_PerfOptimized\_high & ~ & ~ & 46004 & 34161 & 00:08:56 \\ \hline
			~ & impl\_207 & ~ & Performance\_ExtraTimingOpt & -1,31 & 41701 & 34207 & 00:20:42 \\ \hline
			
			synth\_3 & ~ & Flow\_AreaOptimized\_medium & ~ & ~ & 28463 & 23117 & 00:08:11 \\ \hline
			~ & impl\_301 & ~ & Performance\_ExplorePostRoutePhysOpt & -3,39 & 29080 & 23161 & 16:32:38 \\ \hline
			
			synth\_5 & ~ & Flow\_AreaMultThresholdDSP & ~ & ~ & 32517 & 23867 & 00:07:21 \\ \hline
			~ & impl\_506 & ~ & Performance\_Retiming & -1,12 & 31663 & 23913 & 00:11:20 \\ \hline
			
			synth\_7 & ~ & Flow\_RuntimeOptimized & ~ & ~ & 287222 & 214489 & 00:08:18 \\ \hline
			~ & impl\_706 & ~ & Performance\_Retiming & -3,14 & 83566 & 148632 & 00:33:02 \\ \hline
			
			synth\_8 & ~ & Flow\_AlternateRoutability & ~ & ~ & 35422 & 23867 & 00:13:21 \\ \hline
			~ & impl\_801 & ~ & Performance\_ExplorePostRoutePhysOpt & -0,54 & 32879 & 23914 & 00:34:02 \\ \hline
			
			synth\_9 & ~ & Flow\_PerfThresholdCarry & ~ & ~ & 36404 & 28011 & 00:16:26 \\ \hline
			~ & impl\_901 & ~ & Performance\_ExplorePostRoutePhysOpt & -0,42 & 34106 & 28059 & 00:47:13 \\ \hline
			
			synth\_9\_ret & ~ &Flow\_PerfThresholdCarry & ~ & ~ & 36420 & 29278 & 00:8:37 \\ \hline
			
			~ & impl\_901\_ret & Retiming & Performance\_ExplorePostRoutePhysOpt & -0,478 & 33995 & 29324 & 00:24:59 \\ \hline
			
			synth\_9\_flat & ~ & Flow\_PerfThresholdCarry & ~ & ~ & 36404 & 28011 & 00:8:40 \\ \hline
			
			~ & impl\_901\_flat & Full flat & Performance\_ExplorePostRoutePhysOpt & -0,509 & 34185 & 28057 & 00:26:08 \\ \hline
			
			synth\_9\_ret\_flat & ~ & Flow\_PerfThresholdCarry & ~ & ~ & 36420 & 29278 & 00:8:42 \\ \hline

			~ & impl\_901\_ret\_flat & Retiming \& full flat & Performance\_ExplorePostRoutePhysOpt & -0,327 & 34088 & 29324 & 00:24:59 \\ \hline
			
			synth\_9\_incr & ~ & Flow\_PerfThresholdCarry & ~ & ~ & 36420 & 29278 & 00:8:56 \\ \hline
			~ & impl\_9\_incr & & Performance\_ExplorePostRoutePhysOpt & -0,514 & 34191 & 29334 & 00:14:14 \\ \hline
			Non project mode &  &  -directive AlternateRoutability& &  & 36191 & 24397 & 00:06:57 \\\hline
			
			& Non project mode &  &-directive Explore & \textbf{+0.203} & \textbf{32013} & \textbf{23557} & \textbf{00:06:00} \\

			\hline

		\end{tabular}
	\end{table*}

	\subsubsection{Strategies comparison}

	The best configuration is impl\_901, which gave the best WNS of -0.42 ns. It is achieved with the combination of synthesis strategy Flow\_PerfThresholdCarry and implementation strategy Performance\_ExplorePostRoutePhysOpt (table \ref{tab:Best_Implementation_Runs}).

	We can notice from table \ref{tab:Best_Implementation_Runs} that the best slacks, that are -0.42\,ns and -0.54\,ns are obtained for implementations impl\_901 and impl\_801 done with synth\_9 (Flow\_PerfThresholdCarry strategy) and synth\_8 (Flow\_AlternateRoutability strategy,) respectively. 
	This is in accordance with the results obtained in \cite{Lopez}, where the best WNS were obtained 
	among the 8 combinations (table \ref{tab:Run_configurations}) having either Flow\_PerfThresholdCarry or Flow\_AlternateRoutability as synthesis strategy.

	The results are also in good agreement with the implementations runs that
	were performed earlier (in 2025) on another SoC-FPGA circuit \cite{kachkachi_design_2024}  using Vivado 2018.2 version. Indeed, for that project, impl\_93 and impl\_83, that correspond to Flow\_PerfThresholdCarry and Flow\_AlternateRoutability strategies respectively, gave the best results.

	Starting from the best synthesis and implementation configuration, which is synthesis synth\_9 coupled with implementation impl\_901, some further synthesis options (retiming and flatenning) were tested.
	
	Adding the retiming synthesis option "on" didn't enhance timing as showed by comparing the WNS of the runs impl\_901\_ret and impl\_901. The slack in this case is -0.478\;ns which is worse. Nevertheless it brought LUT usage enhancement at the expense of FF usage increase.
	Adding the synthesis flattening option "full" impact is showed by comparing the runs impl\_901\_flat and impl\_901. We can see that full flattening gave worse timing results (slack of -0.509\;ns) with almost same area.
	However, either retiming or full flattening reduced synthesis runtime by 50 \%.
	Moreover, using these options reduced the implementation compile time by more than 50\%.
	
	Finally, the best WNS enhancement was achieved by combining retiming with full flattening. Indeed, the best WNS is obtained in the impl\_901\_ret\_flat, which gave -0.327 \;ns compared to -0.42\;ns of the reference run impl\_901.

	After identifying the best configuration run from strategies exploration using project mode, a fully customized incremental flow and retro-fitting runs were launched using a non-project mode to apply additional fine tunings starting from the best run, i.e., impl\_901\_ret\_flat. A TCL script was written to achieve this incremental and retro-fitting flow. In this flow, an incremental synthesis run was done by performing a pre-synthesis read of a previous synthesis checkpoint with TimingClosure directive. A pre-placement back-annotation was also applied by reading a post-route physical optimization checkpoint. In addition, a post-placement retro-fitting was performed, by reading the previous (reference write\_iphys\_opt\_tcl) post-placement physical optimization script (read\_iphys\_opt\_tcl) just before placement. 
	The incremental flow, that is represented by the impl\_9\_incr run in table \ref{tab:Best_Implementation_Runs}, didn't give the expected results and timing was worse. However, runtime was up to 40\% faster.
	A return to RTL optimization was necessary for the ASIC prototype test-case in order to obtain a closed synthesis timing before moving to implementation stage. Indeed, after RTL enhancement, which consisted in breaking a long data-path to a pipelined structure, a final flow pass in non-project mode without incremental flow neither retrofitting was successfully run. The post-synthesis WNS became +0.346\;ns instead of -1.966\;ns. After placement, WNS was 0.177\;ns and WHS was -0.321\;ns. Post-placement phys-opt was run with -hold\_fix option. The post-place physopt WHS was -0.247\;ns.
	Finally, after routing stage, timing closure was obtained as WNS was +0.203\;ns and WHS was +0.045\;ns. Moreover, ressources utilization was only 32013 for LUT and 23557 compared to previous runs. Implementation runtime was X6 faster, as can be seen from the last line on table \ref{tab:Best_Implementation_Runs} . No further enhancements were needed and post-route phys-opt was not launched.

	\section{Discussion}
	
	As noticed from the \cite{Lopez} study results and from the previous NQR spectrometer SoC-FPGA design, the current test-case confirmed that the best WNS are obtained with implementation that has either Flow\_PerfThresholdCarry or Flow\_AlternateRoutability as synthesis strategies. 
	This demonstrates that synthesis is a crucial step in timing closure. Indeed, as noticed from the executed 260 runs in the current study, implementation strategy doesn't bring expected WNS enhancement if synthesis strategy is not correctly chosen.
	It was showed that global flattening synthesis combined with retiming gave the best timing performance. This is explainable since in full flattening with retiming, the tool has more freedom, typically through cross-boundary optimizations, to achieve better QoR. The main drawback of this approach is runtime. Using incremental synthesis could alleviate to this issue. This synthesis mode is more beneficial at the end of the project when the design is mature. 
	Retiming is very efficient in tackling timing issues but may induce an increase in FF count.
	
	Unexpectedly, incremental flow and retro-fitting didn't enhance the WNS. However, it enabled reducing runtime.
	
	Some ML methods were tried but gave no results for this study. All the conclusions and reflexions presented here are based on classical human observation analysis. 
	AMD intelligent design runs were not possible since the used devices families (Zynq and Kintex) don't support this flow.
	It is worth highlighting that project mode was also good for scripting and enabled taking profit from predefined strategies and settings of the tool. Non-project mode gave more control and fine tuning of the flow.
	
	A good approach to quickly close timing, could be to start with the project mode using either Flow\_PerfThresholdCarry  or Flow\_AlternateRoutability for synthesis combined with impl\_901, impl\_801 or those in table \ref{tab:Run_configurations}. Then switching to non-project mode to manually apply further options settings like retiming and flattening and eventual RTL optimization could bring consistent additional WNS enhancement. This would allow better customization with additional techniques like retro-fitting and eventually working around some bugs or tools issues.

	\section{Conclusion}

	The paper presented the Vivado implementation flow based on UFDM, highlighting different strategies, directives, and practices. Two designs  were used as test-cases for this paper. 
	It was showed that synthesis quality of result plays a crucial role in timing closure. The best two synthesis strategies that were identified are Flow\_PerfThresholdCarry and Flow\_AlternateRoutability. It is a good approach to adopt systematically one of these strategies for synthesis and try to achieve the best WNS at the synthesis stage before moving to the implementation stage.

	\bibliographystyle{ieeetr}
	\bibliography{biblio_arxiv}

\end{document}